\documentclass[12pt,preprint]{aastex}
\usepackage{color}

\newcommand{\change}{ }

\slugcomment{Draft, Astrophysical Journal}

\shorttitle{Time-resolved High-Contrast VLT/SPHERE Imaging of The HR8799 System}
\shortauthors{Apai et al.}

\begin{document}


\title{High-Cadence, High-Contrast Imaging for Exoplanet Mapping: Observations of the HR 8799 Planets with VLT/SPHERE Satellite Spot-Corrected Relative Photometry}


\author{D\'aniel Apai\altaffilmark{1,2,3}}
\affil{Steward Observatory, The University of Arizona, Tucson, AZ 85721, USA}
\email{apai@arizona.edu}

\author{Markus Kasper}
\affil{European Southern Observatory, Garching, Germany}

\author{Andrew Skemer\altaffilmark{3} and Jake R. Hanson}
\affil{Steward Observatory, The University of Arizona, Tucson, AZ 85721, USA}


\author{Anne-Marie Lagrange\altaffilmark{4}}
\affil{Universit\'e Grenoble Alpes, IPAG, 38000 Grenoble, France}

\author{Beth A. Biller}
\affil{Institute for Astronomy, University of Edinburgh, Blackford Hill, Edinburgh EH9 3HJ, UK}

\author{Micka\"el Bonnefoy\altaffilmark{4}}
\affil{Universit\'e Grenoble Alpes, IPAG, 38000 Grenoble, France}

\author{Esther Buenzli}
\affil{Max Planck Institute for Astronomy, K\"onigstuhl 17, Heidelberg, D--69117, Germany}

\author{Arthur Vigan}
\affil{Aix-Marseille Universit\'e, CNRS, Laboratoire d'  Astrophysique de Marseille, UMR 7326, 13388 Marseille, France}

%
\altaffiltext{2}{Lunar and Planetary Laboratory, The University of Arizona, Tucson AZ 85721}
\altaffiltext{3}{Earths in Other Solar Systems Team, NASA Nexus for Exoplanet System Science}
\altaffiltext{4}{CNRS, IPAG, 38000 Grenoble, France}



\begin{abstract}
Time-resolved photometry is an important new probe of the physics of condensate clouds in extrasolar planets and brown dwarfs. Extreme adaptive optics systems can directly image planets, but precise brightness measurements are challenging. We present VLT/SPHERE high-contrast, time-resolved broad H-band near-infrared photometry for four exoplanets in the HR 8799 system, sampling changes from night to night over five nights with relatively short integrations. The photospheres of these four planets are often modeled by patchy clouds and may show large-amplitude rotational brightness modulations. Our observations provide high-quality images of the system. We present a detailed performance analysis of different data analysis approaches to accurately measure the relative brightnesses of the four exoplanets. We explore the information in satellite spots and demonstrate their use  as a proxy for image quality. While the brightness variations of the satellite spots are strongly correlated, we also identify a second-order anti-correlation pattern between the different spots. Our study finds that PCA-based KLIP reduction with satellite spot-modulated artificial planet-injection based photometry (SMAP) leads to a significant ($\sim3\times$) gain in photometric accuracy over standard aperture-based photometry and reaches 0.1 mag per point accuracy for our dataset, the signal-to-noise of which is limited by small field rotation. Relative planet-to-planet photometry can be compared between nights, enabling observations spanning multiple nights to probe variability. Recent high-quality relative H-band photometry of the b-c planet pair agree to about 1\%.
\end{abstract}




\section{Introduction}

Time-resolved precision photometry and spectroscopy of extrasolar planets opens new windows on the atmospheres and origins of these objects: the periodicity of rotational modulations is {\em directly} probing rotational periods; rotationally modulated light curves allow phase mapping of the cloud decks; and {\em light curve evolution} -- changes in the light curves in subsequent rotations -- probes atmospheric dynamics.

For example, recent time-resolved observations of brown dwarfs -- with temperatures and brightness matching those of directly imaged planets -- provided novel insights into the cloud physics and evolution: \citet[][]{Apai2013} used
Hubble Space Telescope (HST) near-infrared (NIR) time-resolved spectroscopy to derive the first cloud maps of brown dwarfs and showed that at the transition between the cloudy L-type and the cloud-free mid-T-type brown dwarfs cloud structures display large cloud thickness variations, but not holes. Based on simultaneous HST and Spitzer observations \citet[][]{Buenzli2012} derived five-layer surface brightness distributions in a late T-type brown dwarf, which revealed a pressure-dependent phase shift corresponding to a large-scale vertically-longitudinally organized atmospheric structure. Space-based surveys by \citet[][]{Buenzli2014} and \citet[][]{Metchev2015} showed that most, if not all, brown dwarfs are variable at wavelengths between 1 and 5~$\mu$m over their rotation period, proving that brown dwarf atmospheres are heterogeneous and suggesting that rotational variability is a powerful tool to study all types of ultracool atmospheres. The reviews by \citet[][]{Metchev2013} and \citet[][]{Burgasser2015} provide more details on other, relevant results from this rapidly growing field. Recently, \citet[][]{Biller2015} presented ground-based precision photometry that led to the detection of rotational modulations in a planetary-mass brown dwarf. Using HST moderate contrast near-infrared observations \citet[][]{Zhou2015} has discovered rotational modulations in the 4-Jupiter-mass close companion 2M1207b, making this the first directly imaged planetary-mass companion with such measurements. These two detections demonstrate the power of rotational modulations in planetary-mass objects.

With the advent of extreme adaptive optics systems (XAO) and the expected bounty of directly imaged exoplanets, accessing the powerful information hidden in temporal modulations -- already studied for brown dwarfs -- requires the ability to obtain {\em time-resolved high-precision photometry in a high-contrast situation}. {\change However, precise photometric calibration of such observations is challenging because it is usually not possible to obtain non-saturated images of the planets {\em and} their host stars or other reference sources; therefore, photometric calibration is not simultaneous and sensitive to temporal variations of the point spread function \citep[][]{Bonnefoy2013}}

\citet[][]{KostovApai2013} used simulated rotational modulations as signal and carried out simulated observations of a hypothetical HR~8799b-like planet for high-quality adaptive optics (AO) system on an 8m telescope, for an XAO system mounted on a 8m telescope, and an XAO system on a 30-m class telescope. Those authors showed that large ($\sim5-10\%$) rotational modulations will be robustly detectable with XAO systems and, in the future, 30m-class telescopes will be able to carry out time-resolved spectral mapping of exoplanets akin to present HST studies of brown dwarfs. {\change \citet[][]{Karalidi2015} showed that high-quality light curves of planets and brown dwarfs can be translated to one- or two-dimensional maps using forward modeling and Markov Chain Monte Carlo-based optimization.} 

In this paper we present results from the first experiment with the SPHERE (Spectro-Polarimetric High-contrast Exoplanet REsearch, \citealt[][]{Beuzit2008}) XAO system mounted on the Very Large Telescope (VLT) aiming to test the feasibility and identify optimal methodology for carrying out time-resolved, relative (planet-to-planet), precision photometry in a high-contrast case. Our observations are based on four short SPHERE/IRDIS datasets on the HR~8799 multi-planet system (discovered by \citealt[][]{Marois2008,Marois2010} and further studied in, e.g. \citealt[][]{Currie2011,Skemer2012}). We demonstrate that data from different nights can be consistently compared and that satellite-spot-modulated artificial planet-injection-based photometry (SMAP) is a capable method that leads to significantly  ($\sim3\times$)  better relative photometric precision than standard photometric methods.

In the following we first describe the observations (\S\,\ref{Observations}), then the different stages of the data reduction and several approaches to measuring the relative photometry of the planets, including the results of our analysis (\S\,\ref{DataReduction}). In \S\,\ref{Discussion} we discuss the relevance of our findings and place them in a broader context. Finally, in \S\,\ref{Conclusions} we summarize the results and the key conclusions of our work.

\section{Observations}
\label{Observations}

The observations were carried out in the Dec 2014 Science Verification run ({ Program ID 60.A-9352(A)}) with the SPHERE { \citep[][]{Beuzit2008} }instrument on {\change VLT's Unit 3 telescope}. We {\change employed classical imaging (CI) mode with apodized-pupil Lyot coronagraphy with} the IRDIS camera \citep[][]{Dohlen2008} in pupil-stabilized mode {\change to allow for angular differential imaging \citep[e.g.][]{Liu2004,Marois2006}}, using {\change the} visible light wavefront sensor. IRDIS is equipped with a Hawaii {\textsc II} 2K$\times$1K detector, which provides an 11\arcsec $\times$ 12.5\arcsec ~field of view and 12.25 milli-arcsecond  (mas) wide pixels.  IRDIS is designed for dual-band imaging (DBI,  \citealt[][]{Vigan2010}), {\change long-slit spectroscopy \citep[][]{Vigan2008}} and polarimetry \citep[][]{Langlois2014}; in our DBI mode the two {\change light beams} passed through the same broadband H-band filter (B$_H$, $\lambda=1.625 \mu$m,	 $\Delta \lambda=0.290 \mu$m) and differed only in their downstream light path through the prism beam-splitter, before being imaged side-by-side on the detector. 

We also utilized SPHERE's ability to introduce satellite spots, by adding a periodic modulation to the deformable mirror { with an amplitude corresponding to 5\% of the full dynamic range of the deformable mirror. These modulations introduced four} satellite spots in a diagonal cross pattern on the detector \citep[][]{Langlois2013}, as shown in Figure~\ref{RawZoom}. In this study we explored the use of these spots for photometric and image quality calibration. {\change No unsaturated image of the star was obtained. }

The data were acquired {\change during } five separate, subsequent nights (see Table~\ref{Log} for details) and {\change consists of a single data cube per night}. The total telescope time for the four identical runs was 3.5 hours. Visual inspection of the frames showed that the data of nights 1, 2, 3, and 5 were of good quality, {\change while the data for night 4} had poor adaptive optics correction due to suboptimal atmospheric conditions. In night 4 less than half of the frames were  useful for our science goals and these did not provide sufficient field rotation. Therefore, {\change data from} night 4 have been excluded from the following analysis.


\begin{deluxetable}{ccccccccccc}
\tabletypesize{\scriptsize}
\tablecaption{Log of the observations. \label{Log}}
\tablewidth{0pt}
\tablehead{
\colhead{Date} & ID &\colhead{MJD} & Rel. Time [h] & \colhead{Field Rot.[$^\circ$]} & Seeing ["] & \colhead{Airmass} & \colhead{Exp. Time [s]} & \colhead{\#Exp.} 
}
\startdata
12/05/2014  & n1 & 56996.00035215  & 0.0~h & 8.7$^\circ$ & 1.00" &1.61-1.51 & 8 s & 218  \\
12/06/2014  & n2 & 56996.99929395& 23.98~h  & 8.7$^\circ$ & 1.05" & 1.62--1.51  & 8 s & 218 \\
12/07/2014  & n3 & 56998.00027928 & 48.01~h &8.5$^\circ$  & 1.35" & 1.64--1.53 & 8 s & 218  \\
12/08/2014  & n4$^{a}$ & 56999.00497827 & 72.12~h &  8.1$^\circ$ & $>1.45"$ & 1.69--1.56  & 8 s & 218 \\
12/09/2014  & n5 & 57000.00740580 & 96.18~h &7.9$^\circ$ & 0.8" &1.73-1.59  & 8 s & 218 \\
\enddata
\tablenotetext{a}{Atmospheric conditions on night 4 were poor; data from these observations were
not included in further analysis. }
\end{deluxetable}

\section{Data Reduction and Results}
\label{DataReduction}

\subsection{Basic Data Reduction}

The {\change files} were downloaded from the ESO archive and each frame in each dataset was inspected {\change by eye}. Frames with low image quality -- typically caused by {\change poor} AO performance -- were excluded from further analysis; these excluded frames usually amounted to about 2-4\% of the data. We reduced the data using a self-developed IDL pipeline.  As first step we identified bad pixels based on their atypical behavior (5$\sigma$ outliers in spatial and temporal distributions in 17$\times$17 pixel search boxes). We identified bad pixels separately for each night's data set and replaced them with the value predicted by spline interpolation of surrounding, valid pixels. This procedure successfully identified and removed almost all bad pixels identifiable by eye.

SPHERE images suffer from slight anamorphic distortion which probably emerges from the common path interface: pixels in y-direction cover slightly smaller (0.6\%) angular scales than in the x-direction. We corrected for the anamorphic distortion by rescaling the image to equal x and y pixel scales using the IDL function {\tt hcongrid}: $y' = y\times1.006$, where $y'$ and $y$ are the corrected and raw y-scales, respectively { following the results of Maire et al. (in press)}.
After this step we separated the left and right {\change IRDIS images} on the detector (both H-band) and reduced them separately, following an identical procedure. The star HR~8799 was occulted by the {\change apodized pupil Lyot coronagraph \citep[][]{Soummer2005}} and partly saturated, making it unsuitable for determining the {\change relative} location of the planets. Therefore, we estimated the location of the star based on the diffraction pattern, specifically, on the location of the four satellite spots. {\change Because the satellite spots have a radially elongated shape we measured the center positions  by identifying the intersection of the vectors connecting and running along the photocenters (spine) of the satellite spots, a technique similar to what is routinely used for aligning Hubble Space Telescope coronagraphic images \citep[e.g.][]{Apai2015}}. 

We then extracted 321$\times$321 pixel (3.92\arcsec $\times$ 3.92\arcsec) sub-images centered on the {\change approximate position of the }star; the {\change sub-}images were {\change then} shifted by a fraction of a pixel to align the center of the stellar residual with the mid-point of the central pixel of the subimage.  

{ The SPHERE files contained the starting and ending parallactic angles, but not the parallactic angles for each individual {\change frame in the data cube}. Given {\change the small field rotations} ($\sim8^\circ$), we calculated the parallactic angles for each frame using a linear interpolation between the starting and ending angles as given by the corresponding POSANG fits header keywords. We used the known constant rotation angle offsets for SPHERE for true north and pupil offset, leading to a 134.1$^\circ$ constant offset correction to transform our interpolated intermediate parallactic angle into the accurate value {\change (Maire et al., submitted)}. }

\subsection{Angular Differential Imaging and KLIP Processing}

\begin{figure}
\epsscale{0.9}
\plotone{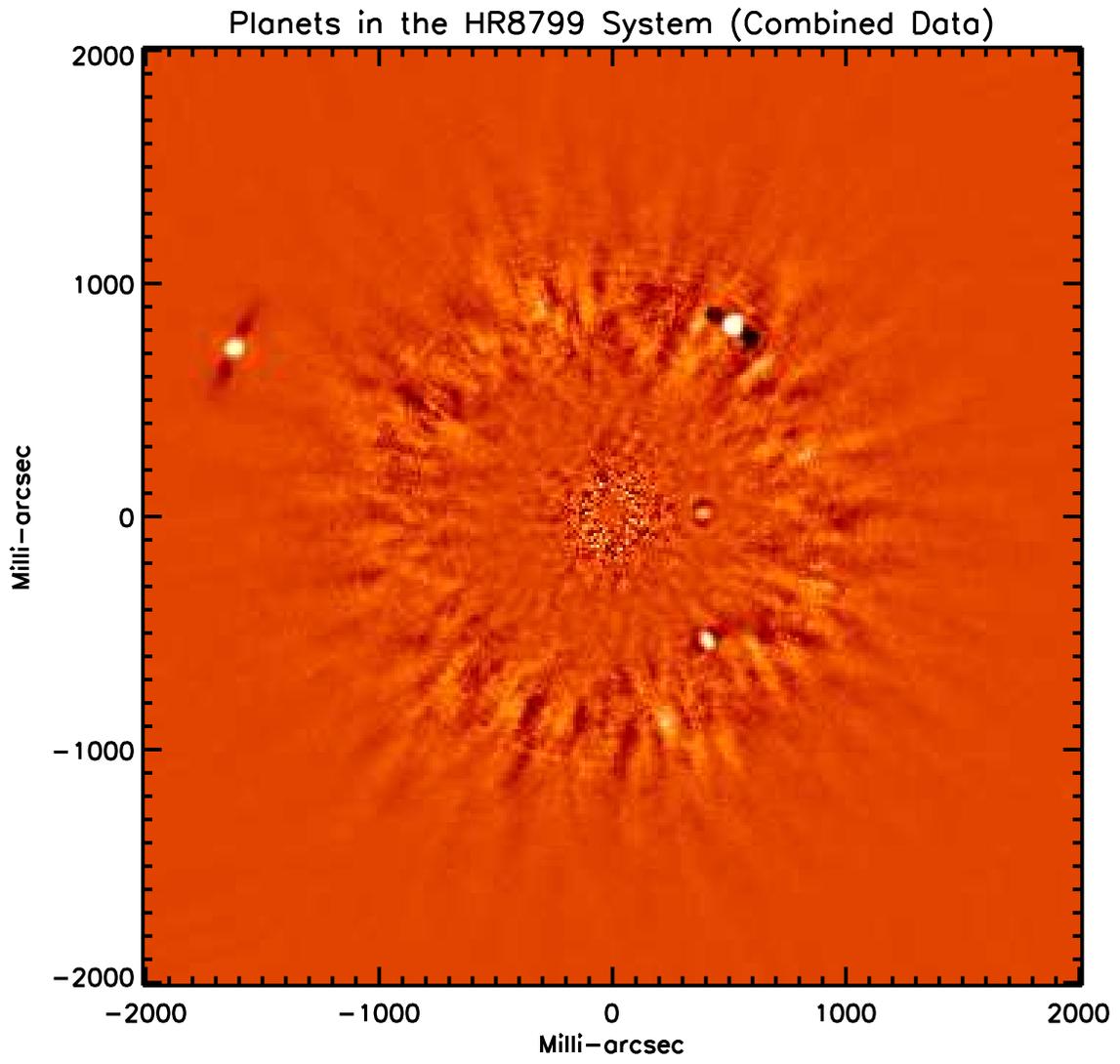}
\caption{Median-combined image from the data from the four nights obtained in our program, logarithmically scaled, from a sector-based KLIP reduction.  \label{Reduced}}
\end{figure}

{\change We followed an optimized principal components (PCA)-based analysis 
to process the cube of sub-images from each night (KLIP: \citealt[][]{Soummer2012}, and \citealt[][]{AmaraQuanz2012})} to identify and remove residuals from the host star's point spread function and associated speckle pattern. Our adaptation of the KLIP algorithm is described in \citet[][]{HansonApai2015}. We excluded from our PCA analysis images with less than 1.5$\times$FWHM-equivalent angular difference from the target image to minimize point source self-subtraction, if at least 5 other suitable frames could be found.
The KLIP-subtracted frames were rotated back to the celestial reference and then median-combined into a single image. This procedure was repeated for the {\change data sets} from each night separately. {\change The median-combination of our four KLIP images is shown in Figure~\ref{Reduced}}.

\begin{figure}
\epsscale{1.2}
\plottwo{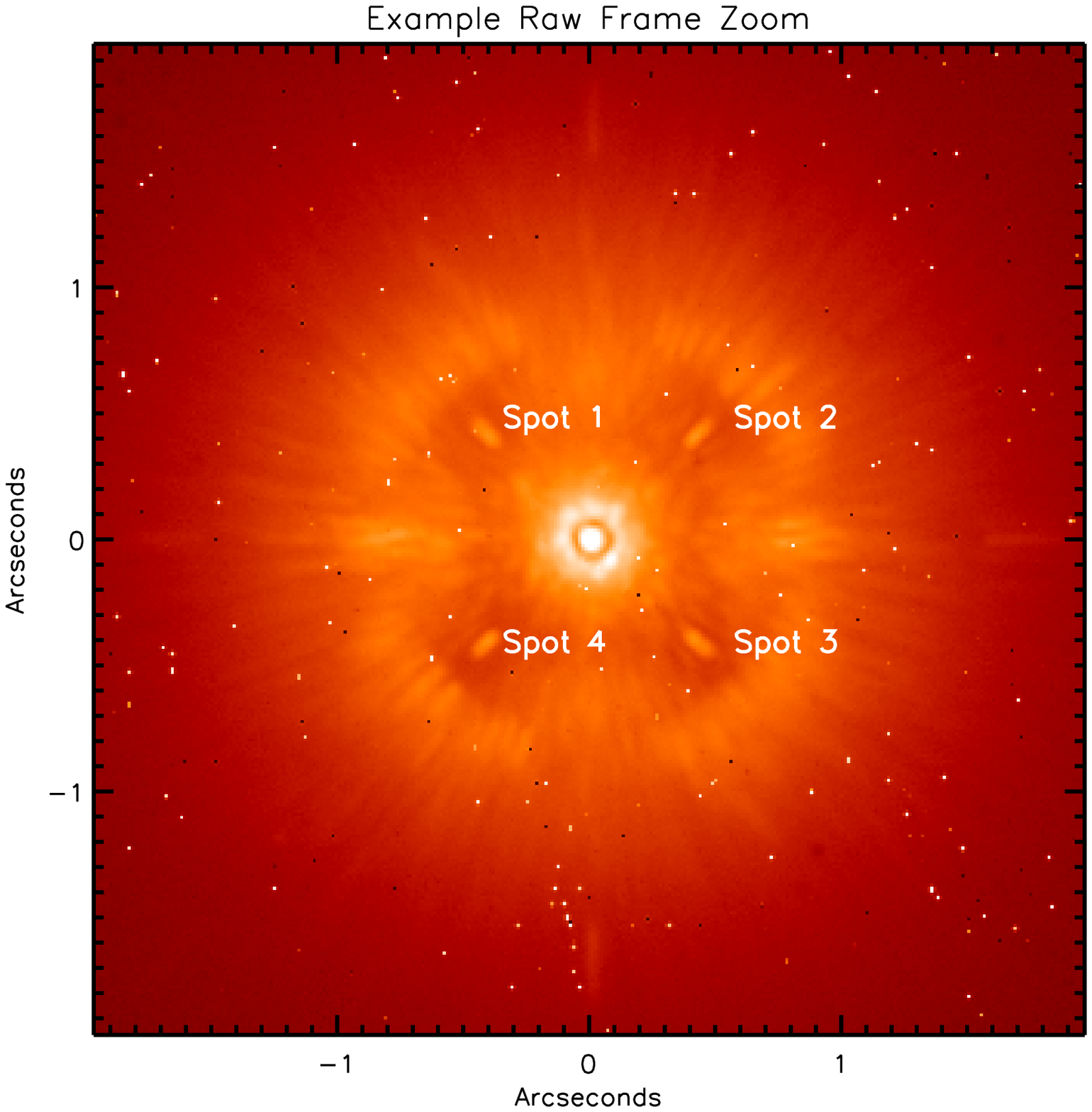}{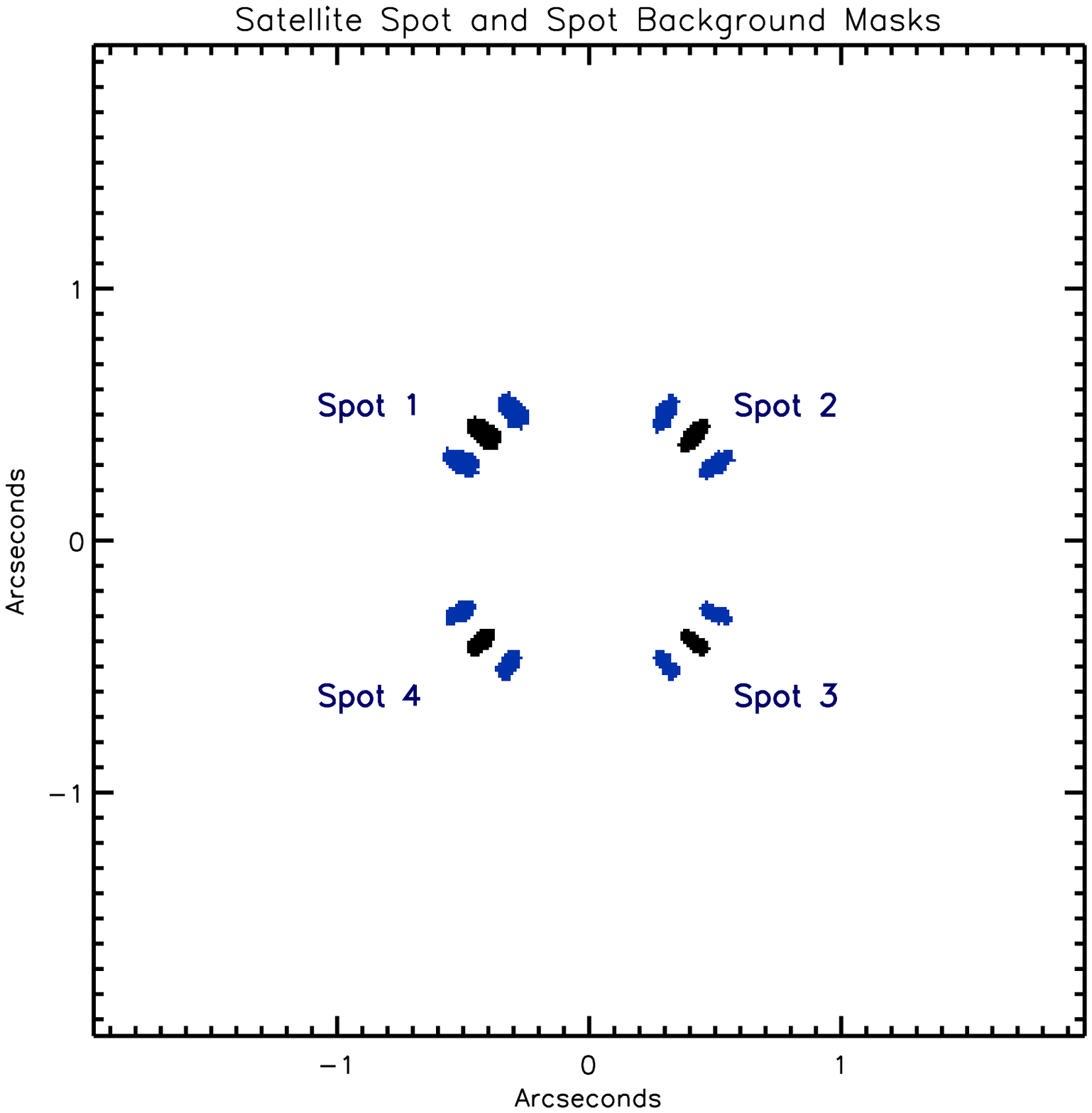}
\caption{{ {\em Left:} Example of a raw sub--image centered on the star, with the four satellite spots labeled, in logarithmic scaling. {\em Right:} The satellite mask (black) and the background mask for the satellite spots (blue) used for measuring the brightness of the satellite spots and their local background.} \label{RawZoom}}
\end{figure}

\subsection{Circular PCA Regions}
\label{CircularPCA}

\begin{figure}
\epsscale{1.2}
\plottwo{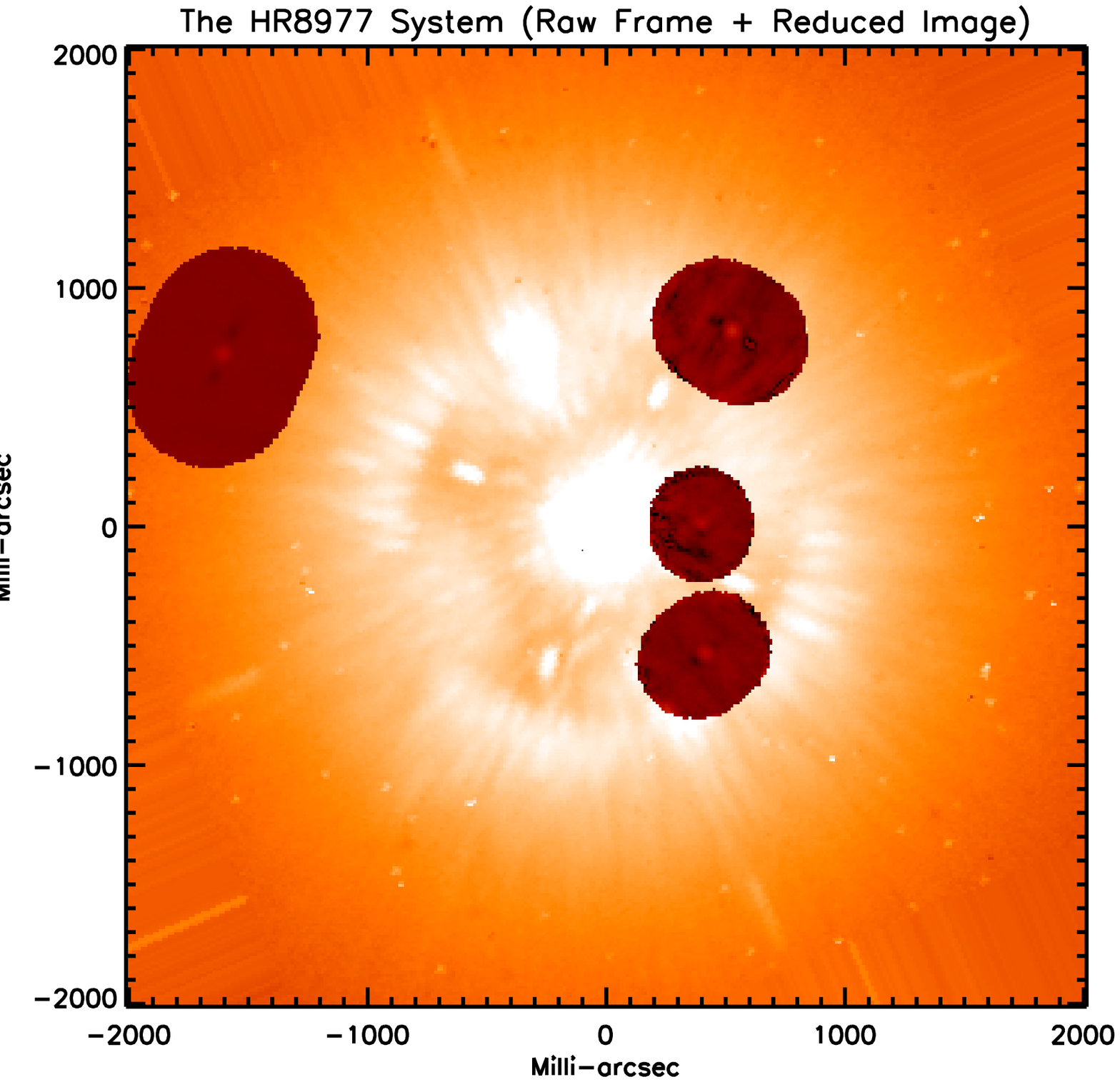}{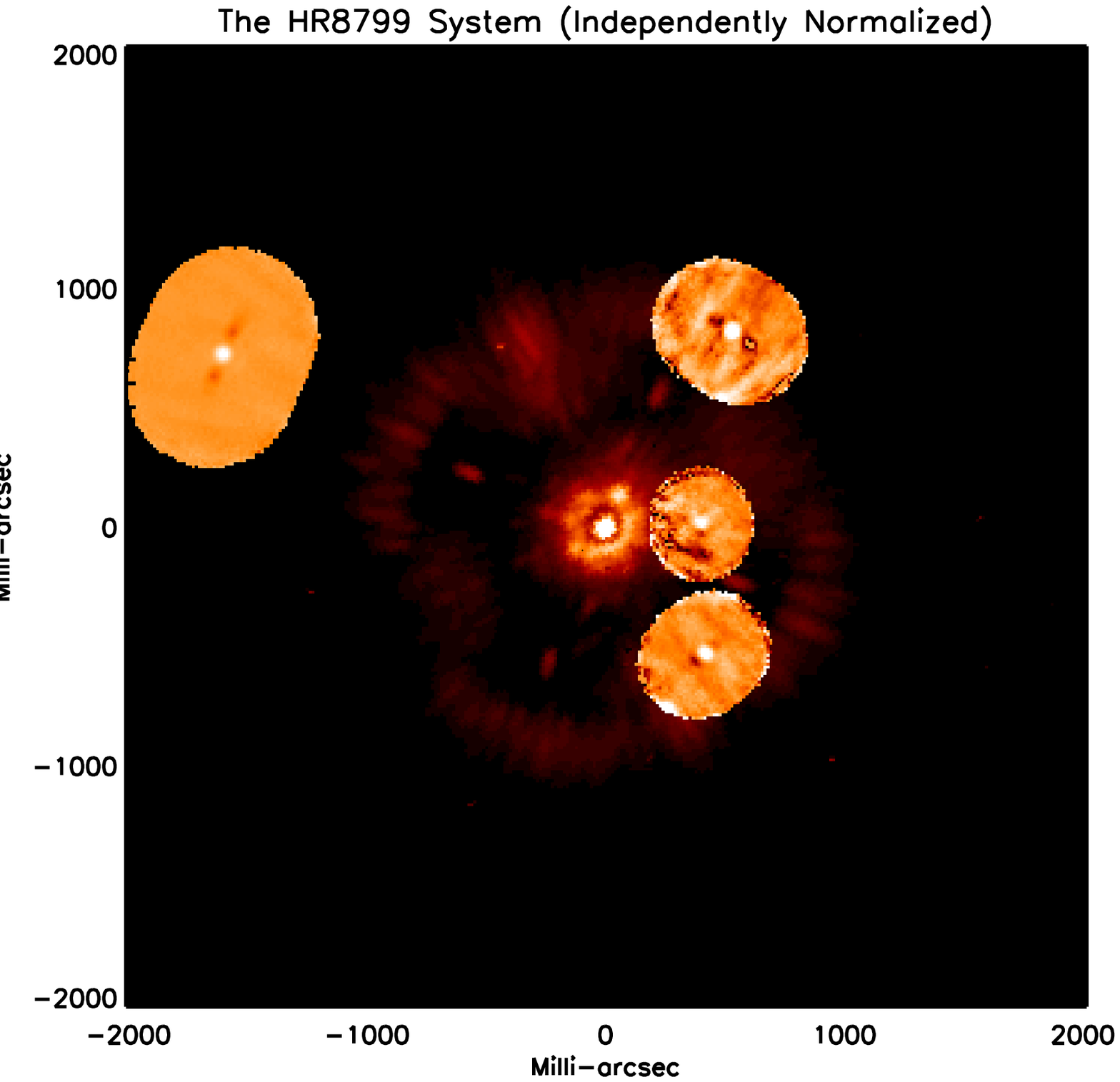}
\caption{{\em Left:} The result of the KLIP reduction of circular patches centered on the four planets, combined with a single raw frame, illustrates the efficient reduction of the stellar point spread function (on logarithmic scale). {\em Right: } The same images, but independently normalized before combining them, to improve the visibility of the image structures in the KLIP-reduced regions. Image is also shown in logarithmic scale.  \label{HyperFinal}}
\end{figure}

The standard approach for PCA-based data reduction is to divide the images in annuli and, within each annulus, in sectors \citep[][]{Lafreniere2007}. Each sector is reduced independently with the KLIP algorithm. While this approach is well-suited for general data reduction problems, such as searches for new planets, in our case it leads to two minor difficulties. First, because we carried out a large number of reductions to optimize the parameters of the {\change algorithm}, the photometric approach, the {\change accuracy of the position and brightness measurements}, etc., the end-to-end KLIP analysis required very significant computing time, even on a multi-core machine. Second, with regularly placed annuli and sectors, sector edges are often close to one of the planets or the varying-sized and shaped apertures we {\change studied aiming to optimize precision of the} planet-to-planet photometry. To circumvent these issues we adopted a slightly different approach: given the known positions of our planets (determined from our images) we defined circular PCA target areas centered on each of the four planets. We only applied the KLIP reduction to these four circular apertures. {\change Figure~\ref{HyperFinal} shows the result of the reduction with the circular KLIP regions.} The circular areas corresponded to only about 5\% of the total {\change sub-image} and by focusing only on these areas, the processing time for an individual dataset was reduced from about 2.5 hours to 3 minutes { on a Mac Pro computer equipped with 2.93 GHz Intel processors}.

\subsection{Preliminary Planet Positions and Planet Point Spread Functions}

As a basis {\change for the optimized procedure for measuring the astrometry and photometry of the planets}, described in the following subsections, we first determined the approximate {\change positions and point spread functions of the planets}. Starting from the {\change pre-processed} image cubes from each night, we { extracted sub-images centered on the {\change left and right beams of the IRDIS detector}, which were reduced independently.  We rotated the sub-images to a north-up position and median-combined the frames in each image.} 

{\change Next,} from each of the eight images (left and right images for each of the four nights) we subtracted their own boxcar-smoothed copy {\change (box width 8 pixels)}, which efficiently reduced the low spatial frequency components in the images. We then combined the eight images into a single, high signal-to-noise image by averaging them and extracted four  {\change 15$\times$15 pixel sized} sub-images approximately centered on planets b, c, d, and e. To determine the positions of the planets within each sub-frame, we cross-correlated the sub-images with a synthetic image consisting of a two-dimensional Gaussian with 5-pixel full-width-at half-maximum. Comparison to the observed planets showed that the synthetic image resembled {\change them} closely and provided a good template for {\change astrometry}. The Gaussian was calculated to be perfectly centered at the {\change mid-point} of the central pixel of the subimage; the cross-correlation determined the relative shift of each planet from the center point of the sub-images. 
By combining the known positions of the four sub-images within the frame and the offsets of the planets within their sub-images, we determined the positions of the planets with sub-pixel accuracy (Table~\ref{Astrometry}). These positions served as {\change initial guesses} for the minimization procedure described below, but their accuracy did not influence our final results. 

Finally, we used the sub-image centered on planet {\em b}, {\change as the highest-quality and cleanest PSF,} in the combined image as a high-fidelity planet template point spread function (pPSF). To remove residual background emission we subtracted the minimum value of the template from this subimage and then masked out low-level residual starlight in the corners of the template. 
At the end of this process, thus, we obtained preliminary positions and point spread functions for the planets, which were refined in subsequent steps.

\subsection{Artificial Planet Injection and Residual Fits}
\label{ArtificialPlanetInjection}

We determined the position and brightness of each planet in each of the four datasets (nights 1, 2, 3, 
5) by injecting {\em artificial} planets { (PSFs derived from planet b)} with negative count rates and positions approximately matching those of the real planets (determined from non-PCA-reduced images), and then by minimizing the subtraction residuals as a function of position and flux of the simulated planets (e.g. \citealt[][]{Lagrange2010,Marois2010SPIE, Bonnefoy2011})

1) In order to estimate the { brightnesses of the planets in our data, we added negatively scaled copies of pPSFs to individual raw data frames}. The pPSFs were added in a super-sampled space: both the target image and the pPSF were interpolated to a ten times finer pixel grid before summing them and sampling the results back to the original resolution. The pPSFs were scaled by their total counts.

2) We repeatedly reduced the images using different pPSF scaling and different position angles with the goal to identify the position and brightness that minimize the residuals. We measured the square of the residuals as a function of position and brightness. The squared residuals were integrated in apertures of different shapes (see Section~\ref{ArcApertures}) placed at the approximate positions of the planets.

3) To determine the best-fit position and brightness of each planet we fitted parabolas on the residuals as a function of the two variables (position angle and brightness; see \S\,\ref{SectionAstrometry} for the choice of parameters). The data points were weighted inversely with the sum of the squared residuals, i.e. values giving better subtraction had a higher weight in the fit. 

\begin{figure}
\epsscale{0.9}
\plotone{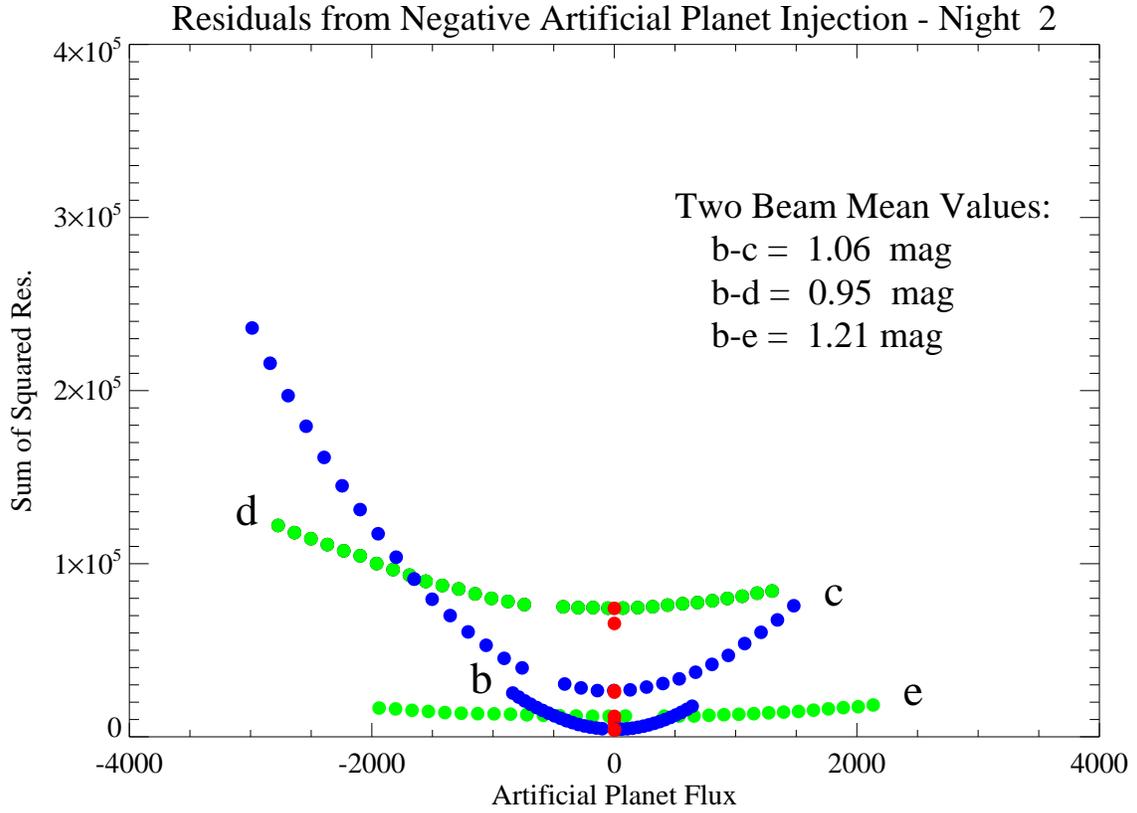}
\caption{An example result from the artificial planet injection photometry for one of the two images obtained in night 2: Sum of the squared residuals as a function of artificial planet flux relative to the best-fit flux. Blue symbols are measurements. Red symbols mark the minima predicted by parabolic fits for {\em both images} (i.e., left and right beams, two red symbols) for each planet and are in excellent agreement. The corresponding plots with results for the other three nights and for the position determination are very similar and not shown here but results are captured in Tables~\ref{Astrometry} and \ref{Photometry1}.  \label{ParabolicFit}}
\end{figure}

Figure~\ref{ParabolicFit} shows the sum of the squared residuals as a function of brightness of the planets and the best-fit parabola for each { planet in Night 2}. Table~\ref{Astrometry} summarizes the positions of the four planets determined through the non-PCA process and through the parabolic fits. 

The planet position and brightness optimization procedure required a few thousand independent data reductions.


\subsection{Astrometry}
\label{SectionAstrometry}
{ When data with very limited field rotation is combined, angular differential imaging can lead to arc-like oversubtraction artifacts due to partially overlapping and therefore self-subtracting images of the planets.} While it is relatively straightforward to determine the planet-star separation in our images, determining the precise position angles of the planets is somewhat more challenging due to the arc-like self-subtraction resulting from the small field rotations present in our images.
{ Therefore, we derived the position angles of the planets using the artificial planet injection method described in \S,\ref{ArtificialPlanetInjection}. }

The planet-star projected separations on the detector are well-determined by our high signal-to-noise non-PCA-reduced images. {\change Based on  the manual inspection of the image quality and the subtraction residuals we estimate that our measurements of the planet-star separations are accurate to about 0.5~pixels or about 7~mas.
The astrometric measurements are shown in Table~\ref{Astrometry}.}

\begin{deluxetable}{cccccc}
\tabletypesize{\scriptsize}
\tablecaption{Planet astrometry. \label{Astrometry}}
\tablewidth{0pt}
\tablehead{
\colhead{Parameter} &\colhead{Night} &\colhead{Planet b} & \colhead{Planet c} &  \colhead{Planet d} & \colhead{Planet e}\\}
\startdata
$\rho$ &  all, non-pca   & 1\farcs729"$\pm$0\farcs007 & 0\farcs947$\pm$0\farcs007 & 0\farcs654$\pm$0\farcs007  &  0\farcs381$\pm$0\farcs007  \\ 
\hline
$\phi$ &  n1        & 65.93$^\circ$ & 328.25$^\circ$   & 216.85$^\circ$ & 272.48$^\circ$  \\
$\phi$ &  n2    & 66.14$^\circ$ & 327.82$^\circ$ &  216.36$^\circ$ & 271.35$^\circ$ \\
$\phi$ &   n3   & 65.82$^\circ$ & 327.83$^\circ$ &  217.66$^\circ$ & 272.35$^\circ$ \\
$\phi$ &   n5   &  65.82$^\circ$    &  327.18$^\circ$   & 218.16$^\circ$ & 273.08$^\circ$ \\
\hline
$\phi$ &  Median         &    65.93$^\circ$  &   327.83$^\circ$  &    217.66$^\circ$  & 272.48$^\circ$\\
$\phi$ &  Mean           &    65.93$^\circ$  &   327.77$^\circ$ &  217.26$^\circ$   & 272.32$^\circ$  \\
$\phi$ &  Std. Dev.     &    0.15$^\circ$  &   0.44$^\circ$ &  0.81$^\circ$   & 0.72$^\circ$  \\

\enddata
\end{deluxetable}

\subsection{Satellite-spot Behavior}
\label{SatSpotSection}

Our images included four satellite spots, injected {\change using the deformable mirror of SPHERE as} references for image quality and flux.  The satellite spots -- as part of the stellar PSF -- are efficiently removed by the KLIP processing; therefore, we carried out the satellite spot photometry on each frame of the {\change pre-processed} data, i.e. before KLIP processing. { The satellite spots have an irregular  shape, i.e. not well matched to a circular or elliptical mask. Therefore, we created a binary mask with values 1 at pixels corresponding to the satellite spots and values of 0 at all other positions. 
The transparent pixels (1) in the masks were identified by applying an intensity cut to one of our SPHERE sub-images, adjusting cut levels to match the level where the satellite spot edges emerge from the background levels. Finally we eliminated other structures outside the satellite spots by manually setting pixel values to zero. This mask has been carefully inspected by eye and compared to the actual images to ensure that it captures the satellite spots; however, we note that in this study we only use {\em changes} in the satellite spots, therefore our study is not sensitive to or dependent on precisely determining the edges of the spots.
The right panel of Figure~\ref{RawZoom} shows the satellite spot mask and the similarly derived satellite spot background mask (described below). On each image we determined the total flux in the four satellite spots in the mask apertures.

  We considered the question whether the satellite spots should be background-subtracted. Such subtraction is a correct step if the satellite spots sit on top of an unrelated (i.e. non-correlated) background or if the satellite spots sit on a background that is anti-correlated with the spots. The latter behavior is expected, for example, if the background levels below the satellite spots are dominated by uncorrected stellar halo, which has lower values on higher quality images and higher values on lower quality images.

To assess the background behavior {\change we measured its level near the location of the satellite spots using a mask-based approach}. The background mask was defined by combining two copies of the satellite spot mask rotated by $\pm14^\circ$ round the star's position as pivot point.  {\change The angle was determined by eye to place the background apertures at locations that are as free as possible from PSF structures. The two background spots were positioned symmetrically} left and right from each satellite spot (see Fig.~\ref{RawZoom}). We measured the satellite spot background levels in the mask apertures, analogous to the satellite spot brightness measurements.

Our analysis showed that the background levels are very highly {\em positively correlated} with the satellite spots, showing that the background flux is dominated by flux that directly scales with the image quality. Therefore, subtracting the background levels from the spots would reduce actual correlated signal and would be the incorrect step. Consequently, we did not apply background correction to the spots.}

\begin{figure}
\epsscale{0.9}
\plotone{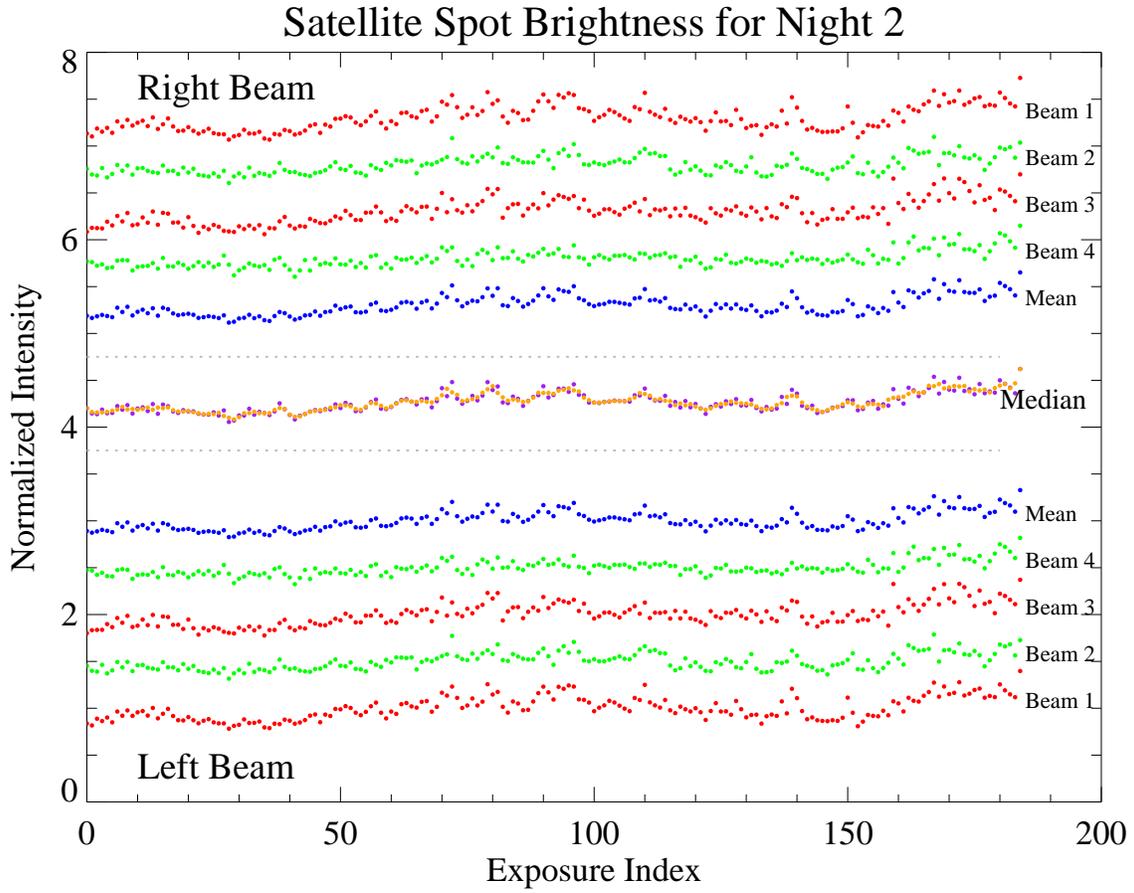}
\caption{Normalized intensity for the four spots (red and green) and their mean value (blue) for both left channel (bottom panel) and right channel (top panel) from the data acquired during night 2. The median curve (purple) as well as the box-car smoothed median (orange) derived considering all the eight spots are shown (middle panel). The curves are vertically shifted for clarity. All curves exhibit highly correlated temporal variability. \label{SatSpots}}
\end{figure}

The satellite spots display temporal variations from frame to frame.  As an example Figure~\ref{SatSpots} shows the photometry for the left and right images, respectively, in our night 2 dataset. We normalized the photometry of each spot by dividing it with its time-averaged value. 
The flux of the individual spots is highly variable: the 1$\sigma$ standard deviation of the measured fluxes is 8--12\%. The mean value of the four spots in each of the right and left images (also shown in Figure~\ref{SatSpots}) shows a standard deviation of $\sim$10\%.
As a test of how well the fluxes in individual satellite spots are influenced by shared variations between all four spots or by individual effects, we divided each satellite spot value by the mean value of the four spots. These individual normalized satellite spots show smaller, but still significant variations: the typical standard deviation is 3.6--4.3\%. This corresponds to about 40\% of the total variations; therefore, we conclude that the shared variations and individual variations contribute about evenly to the temporal behavior of the satellite spot fluxes. 



We explored the correlations between the individual satellite spots and their mean using the 
Spearman's correlation test. Spearman's $\rho$-test is a non-parametric test that is based on
the ranking of the data points rather than their values, which makes it more robust than the simple linear correlation test for evaluating the presence of correlations between quantities that may have different underlying distributions \citep[][]{Press2002}. $\rho$ values close to $+$1 represent very strong correlation; positive values close to zero represent weak or no correlation; values close to $-$1 mean strong anti-correlation. We found that the typical $\rho$ values between spots on the opposite sides of the star (1 and 3, and 2 and 4, Fig~\ref{RawZoom}) are in the 0.78--0.88 range, while adjacent spots show a weaker correlation ($\rho$=0.72-0.82).  Typically, the Spearman $\rho$ coefficients for individual spots range between 0.83--0.96, with most values above 0.90, indicating very strong correlation.

We also found that the correlation coefficients also reflect the overall data quality: we found stronger correlations between values in nights 2 and 3, for which the observing conditions were good, and weaker correlations between the values measured for nights 1 and 5, for which the conditions were less stable.

\begin{figure}
\epsscale{1.0}
\plottwo{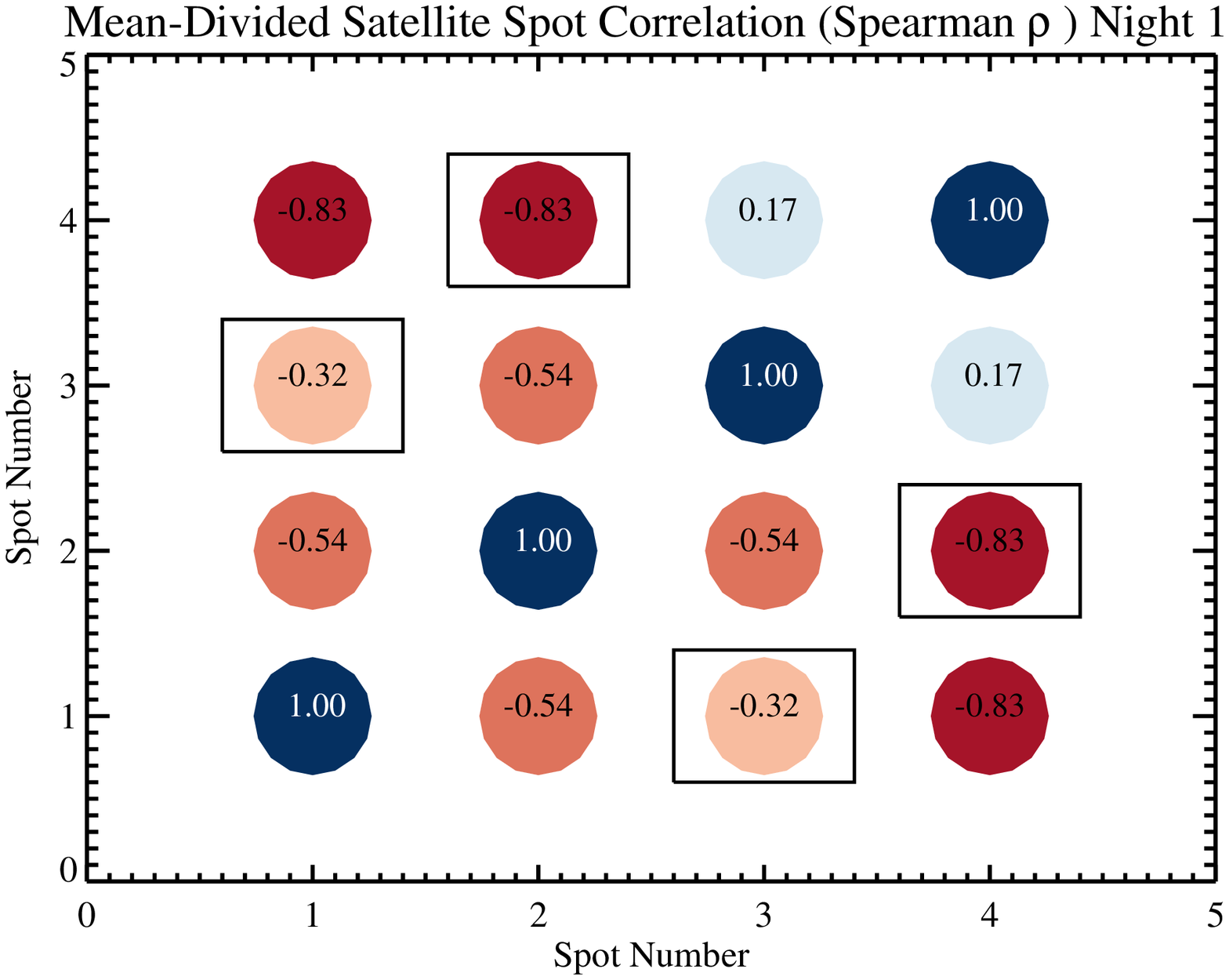}{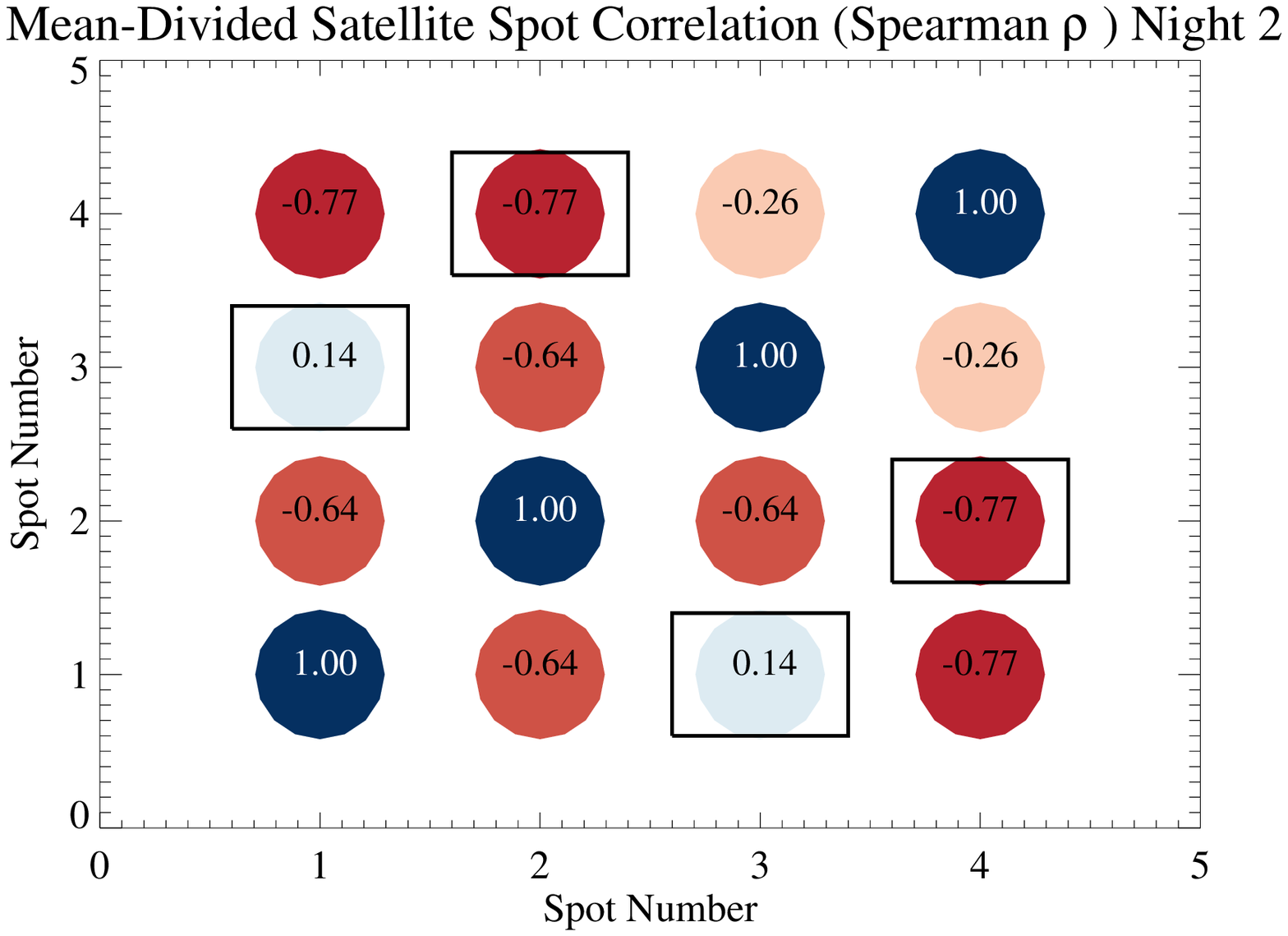}
\plottwo{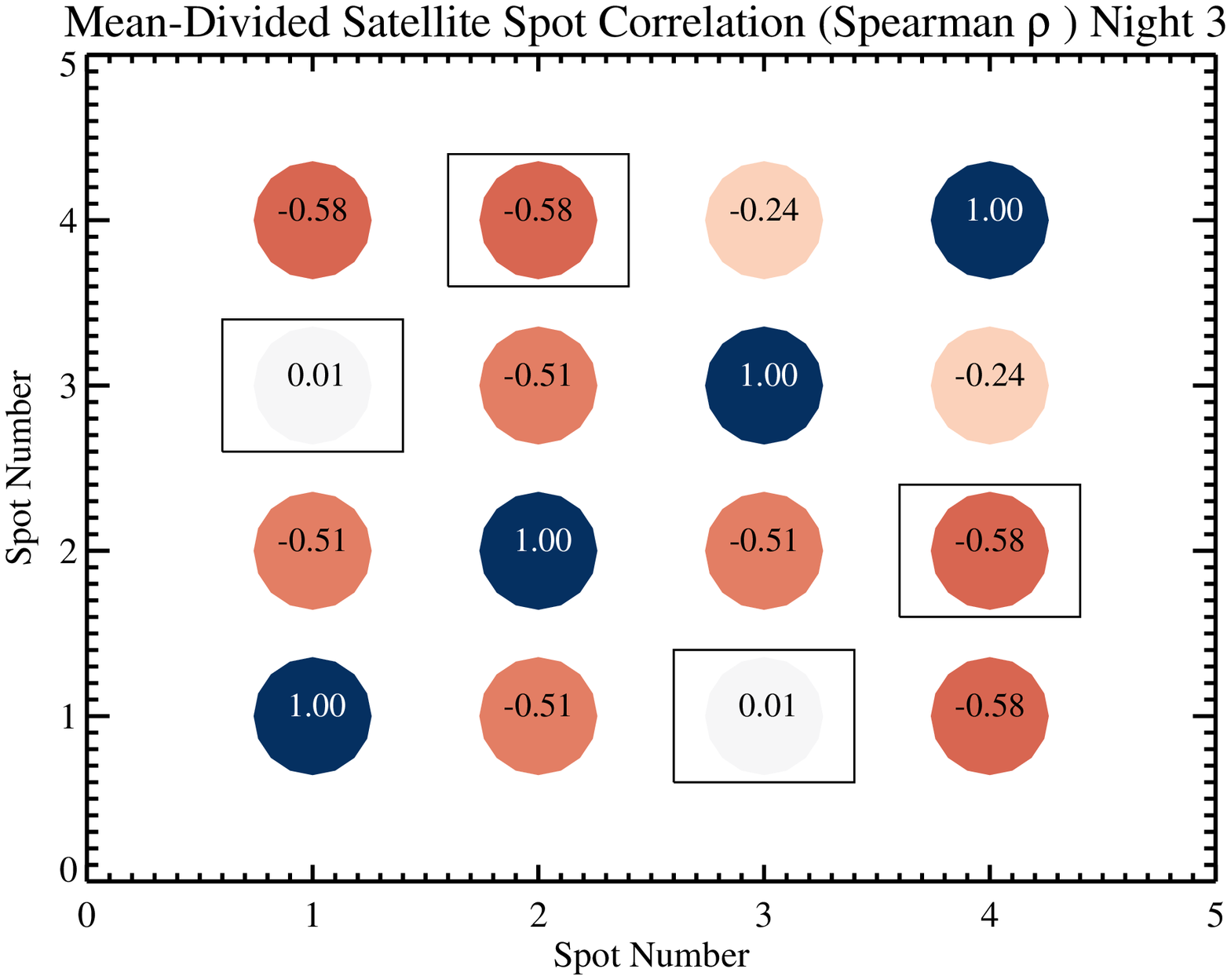}{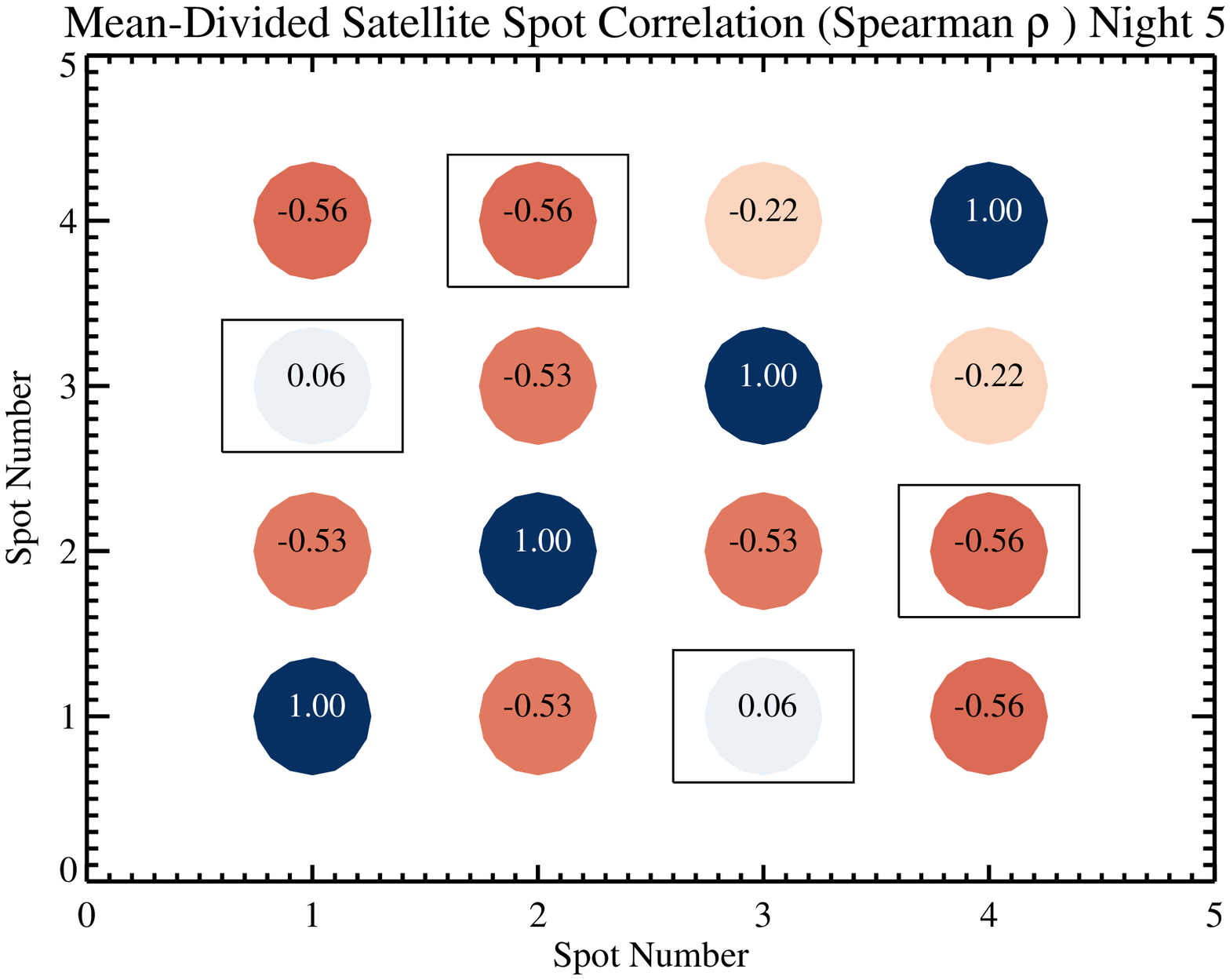}
\caption{Spearman $\rho$ correlation values for the four satellite spot lightcurves {\em after} dividing them with the frame-by-frame mean value, removing the dominant shared variations. The symbols in rectangular frame are satellite spot pairs on opposite sides of the star, whereas the other symbols mark adjacent spots or self-correlation.
Note, that the overall correlation pattern indicates strong anti-correlation between some spots (e.g. 2 and 4) and that the pattern is very similar for all four nights. \label{SatSpotsCorrelation}}
\end{figure}

The frame-to-frame mean value of the satellite spot brightnesses captures the dominant shared variations. It is instructive, however, to explore the satellite spot behavior beyond the variations they share. We divided each satellite spot with the mean value of the satellite spots on the given image, thus removing shared variations. This step, as explained above, reduces the scatter in the satellite spot light curves by about a factor of two. Even cursory inspection of the corrected lightcurves reveals that the remaining scatter is mostly not white noise (Fig.~\ref{SatSpots}). In Figure~\ref{SatSpotsCorrelation} we show four diagrams that illustrate the Spearman $\rho$ correlation coefficients between pairs of spots {\em after} subtraction of their correlated variations.

These figures reveal two surprising results: First, the spots show primarily no/weak anti-correlation or, in several cases, relatively strong anti-correlation (e.g. spots 2 and 4). Second, the anti-correlation pattern is remarkably constant over the four nights: while opposite spots 1 and 3 are not correlated; opposite spots 2 and 4 are always strongly anti-correlated. Adjacent spot pairs 2--3, and 1--4 are also always anti-correlated.
 
Next we carried out principal components analysis (PCA) of the eight time-variable satellite spots (normalized to the mean intensity) to better characterize their behavior. The PCA results have, again, been remarkably consistent over all four nights: the primary principal component accounts for 47--59\% of the variance, the secondary principal component accounts for 21--32\%, the tertiary component accounts for 9--19\%; further components are all below 0.1\%. These results suggest that -- in addition to the mean variation -- always three components dominate the variations in the four satellite spots, with one alone accounting for about half the variations. 
{ The fact that the overall satellite spot time evolution follows a strict pattern suggests that, using longer datasets, a quantitative model could be constructed to describe the variations of the different components. We propose that such a quantitative model can be used to reach further significant improvements in the planet-to-planet photometry by allowing a determination of the effective image quality across the image.}

Based on these findings we conclude that: a) all spots are strongly correlated; b) spots on opposite sides of the star are more strongly correlated than spots adjacent to each other; c) all spots are strongly correlated with the mean spot value; d) about half of the observed variations in the satellite spots are shared between the four spots and are well captured by their mean value; d) even after normalization by the mean spot value significant anti-correlation patterns are present that are repeatable for all four nights; and, e) the remaining variations can be described to $>$99\% by three principal components, the most important of which accounts for about half of the variance. 

\subsection{Planet-to-planet relative photometry}



To measure the brightness of the planets we started from the same artificial planet injection-based procedure as used for the precision astrometry (\S\ref{SectionAstrometry}). However, unlike in the more straightforward astrometry process, we tested a variety of approaches to identify the most accurate procedure for the photometry. Given that we only had four data points {\change for each night}, our primary goal here was to evaluate the optimal photometric approach. The fact that our datasets consist of four planets measured in four nights, i.e. 16 photometric points, allowed us to test and compare the precision reachable with the different approaches. We did not attempt to derive absolute planet brightnesses, given the lack of suitable calibration observations, but instead focused on determining the relative brightnesses of the planet pairs ({\em b-c, b-d, b-e}) for each of the four nights. 

For the purposes of comparing the different photometric approaches we adopted the assumption that all planets are of constant brightness, i.e. the level of brightness differences measured between any two planets is dominated by our photometric uncertainty. With this assumption the {\change standard deviations of the measurements for the four } planet pairs provide a good basis for comparing the relative accuracy of different photometric approaches.

In the following we discuss the different parameters tested and our findings. Table~\ref{Photometry1} summarizes the results of the photometry with different approaches.


{\bf PCA Target Sectors vs. Circular PCA Target Regions:} We repeatedly re-reduced all data with an  annular sector-based KLIP routine and with a KLIP routine that focused on four circular patches centered at the known positions of the four planets. We found that the circular patches-based KLIP analysis significantly improved the accuracy of the photometry from $\sim$0.20 mag to $\sim$0.10 mag. {\change We attribute this improvement to gains due to the lack of edge-effects near the planets and the fact that the circular regions provide correction centered on the planets, while the sector-based reduction does not.}

{\bf Square vs. Arc-like Apertures for Residual Measurements}
\label{ArcApertures}
In evaluating the residuals from the negative planet injection we measured the squared residuals in a small 7 pixel by 7 pixel square centered on the planet; then we re-reduced the data with the arc-like (angular) apertures. {\change The angular size of the arcs were determined to optimally cover the residuals on the image and were prescribed as $6^\circ + 9$ pixels. The apertures had fixed radial widths (13 pixels=159 mas or approximately 4 FWHMs)}. This definition resulted in arc-like apertures that fully covered the planet's trace on the images, considering both the diffraction-limited resolution of the images and the field rotation. 
{\change The comparison of the photometric accuracy results, as discussed later, showed that adopting the arc-like apertures led to significant improvement over the square apertures (see Table~\ref{Photometry1}).}

{\bf SMAP: Satellite-spot Modulated Artificial Planet Injection-based Photometry:} We also reduced our data following three different methods: a) ignoring the information in the satellite spots; b) by modulating the injected planets' brightness in {\em each raw} image by the normalized mean flux of the satellite spots measured in the same image; and, c) by { modulating the planets with a 2-element sliding temporal average of the median value of the four spots}. 

While case {\em a} provided the baseline reference, case {\em b} corresponded to the hypothetical situation in which the frame-to-frame variations of the satellite spots accurately reflect the variations in the combined AO correction and sky transmission, while {\em c} corresponded to the hypothetical situation where the satellite spot measurements are "noisy" and the key information they carry is the overall shape of the correlated component.

Comparison of the photometry achieved in the three cases demonstrated that the satellite-spot modulated artificial planets-based photometry represents a clear and significant improvement over the unmodulated artificial planet-based photometry; and that case {\em b} also provides better photometric accuracy than case {\em c}, i.e. frame-to-frame variations in image quality are important and can be partly corrected by the mean of the satellite spots.

\begin{deluxetable}{lcccccc}
\tabletypesize{\scriptsize}
\tablecaption{Relative planet photometry from angular segments-based PCA reduction. \label{Photometry1}}
\tablewidth{0pt}
\tablehead{
\colhead{Night} & PCA   &  Residual  & Modulation by      &\colhead{b-c} &\colhead{b-d} &\colhead{b-e}  \\
\colhead{  }       &            &  Aperture  & Satellite Spot  & \colhead{[mag]} &\colhead{[mag]} &\colhead{[mag]}  }
\startdata
1 & Angular & Square & No                  & 1.52 & 1.30   &1.65   \\ 
2 & Angular & Square & No                 & 1.06 & 1.04   & 1.10      \\ 
3 & Angular & Square & No                 & 0.97 &  0.79  & 1.10     \\ 
5 & Angular & Square & No                 & 1.35 & 0.97 &1.43     \\ 
\hline
Mean         			& Angular & Square & No                 &  1.22 &  1.03 &  1.32    \\
Std. Dev.  			& Angular & Square & No                  &  0.25       &  0.21 & 0.27     \\
\hline
\hline
1  & Angular & Square & Mean    & 1.52   &  1.30 & 1.65 \\  
2  & Angular & Square & Mean   & 1.08 & 1.05 & 1.13 \\
3  & Angular & Square & Mean   & 1.00 & 0.68 & 1.06 \\
5  & Angular & Square & Mean   & 1.18  & 0.72 & 1.14  \\
\hline
Mean & Angular & Square & Mean   & 1.20   &   0.94 &    1.25         \\
Std. Dev. & Angular & Square & Mean   & 0.23  &   0.29  & 0.27    \\
\hline
\hline
1 & Angular & Arc-shaped & Mean    & 1.46   &  1.22 & 1.61  \\  
2 & Angular & Arc-shaped & Mean   & 1.08& 1.08 & 1.17  \\
3 & Angular & Arc-shaped & Mean  & 1.07 & 0.77 & 1.02  \\
5 & Angular & Arc-shaped & Mean  & 1.18 & 0.62 & 1.14   \\
\hline
Median & Angular & Arc-shaped & Mean    & 1.18  &   1.08 &    1.17          \\
Std. Dev. & Angular & Arc-shaped & Mean & 0.18   &   0.135  &   0.26     \\
\hline
\hline
1 & Angular & Arc-shaped & No  & 1.46   &  1.22 & 1.61  \\  
2 & Angular & Arc-shaped & No  & 1.06& 1.07 & 1.13  \\
3 & Angular & Arc-shaped & No & 1.04 & 0.85 & 1.11  \\
5 & Angular & Arc-shaped & No & 1.28 & 0.96 & 1.30   \\
\hline
Median  & Angular & Arc-shaped & No   & 1.28   &   1.07 &    1.30        \\
Std. Dev. & Angular & Arc-shaped & No  & 0.20   &   0.16  &   0.23  \\
\hline
\hline
1 & Circular & Arc-shaped& Smoothed Median & 1.07   &  0.83 & 1.23  \\  
2 & Circular & Arc-shaped & Smoothed  Median & 1.11  &1.01 & 1.17    \\
3 & Circular & Arc-shaped&  Smoothed Median &  0.78 & 1.07 & 1.37   \\
5 & Circular &Arc-shaped &  Smoothed Median &  0.96 & 1.14 & 1.32  \\
\hline
Mean & Circular & Arc-shaped &  Smoothed Median     & 0.98   &   0.99 &    1.27    \\
Std. Dev. & Circular & Arc-shaped &  Smoothed Median   & 0.15   &  0.13  & 0.09     \\
\hline
\hline
1 & Circular &Arc-shaped& Mean  & 1.11   &  0.91 & 1.34  \\
2 & Circular & Arc-shaped & Mean   & 1.06  & 0.95 & 1.21   \\
3 & Circular & Arc-shaped& Mean  & 0.92 & 0.94 & 1.22   \\
5 & Circular & Arc-shaped& Mean   & 0.95 & 1.15 & 1.34   \\
\hline
{ Mean} & Circular & Arc-shaped & Mean   &{ 1.01  } & {  0.99 }& { 1.27 } \\
{ Std. Dev.} & Circular & Arc-shaped & Mean & { 0.09 }  & {  0.11 } & {   0.07 } \\
\hline

\enddata
\end{deluxetable}

\section{ Discussion}
\label{Discussion}

The analysis presented in the previous section, although based on a relatively small data set obtained at high airmasses, provides interesting insights into six topics: the planet's positions, rotational variability in the four planets, the behavior and best use of satellite spots for relative photometry, the optimal approach for precision relative photometry, the ability to combine data from different nights, and potential for exoplanet mapping with the precision demonstrated here. In the following we briefly discuss each of these six topics.

\subsection{Planet Positions}
By injecting artificial planets in the individual raw frames at varying positions and then fitting a parabola on the residuals measured in the final, reduced frame in arc-like apertures, we determined the positions of the four planets separately for each of the four nights. The median-combined positions for the four planets
are: {\em b}: $\rho=1\farcs729\pm0\farcs007$ and $\phi=65.93^\circ\pm0.15^\circ$;   {\em c}: $\rho=0\farcs947\pm0\farcs007$ and $\phi=327.83^\circ\pm0.44^\circ$; 
{\em d}: $\rho=0\farcs654\pm0\farcs007$ and $\phi=217.66^\circ\pm0.81^\circ$; {\em e}: $\rho= 0\farcs381\pm0\farcs007$ and $\phi=272.48^\circ\pm0.72^\circ$.
We note that these positions are consistent with the latest published positions of the four planets and orbital fits \citep[e.g.][]{Maire2015,Pueyo2015}, but given the focus of the paper and fact that our observations only contribute a single new epoch to the existing astrometric data on the planets, we do not attempt to derive revised planetary orbital elements.

\subsection{Satellite Spot Behavior}

We used four satellite spots, introduced via modulating the instrument's deformable mirror, as probes of image quality. We found that each of the satellite spots varies significantly (standard deviation $>$1$\sigma$), partly reflecting shared variations and partly due to individual variations. We found that the mean spot brightness is a good proxy of the image quality (presumably including variations in the AO correction and sky transparency) and that frame-to-frame changes in this value are real and carry useful information for the data reduction. We found that spots on opposite sides of the star, introduced by the same deformable mirror modulation, show slightly higher correlation than adjacent spots, based on a Spearman non-parametric rank-test.

The fact that the spots show $\sim$4--5\% level individual variations demonstrates that no single spot or simple combination of spots can be used as a perfect image-to-image flux reference; nevertheless, we found that the mean flux of the four satellite spots provides a useful proxy for the general image quality variations. Artificial planet injection-based photometry, when modulated with the mean of the satellite spots, provided the most accurate photometry in our study, demonstrating the value of the satellite spots.

{ The satellite spots share (correlated) variations that account for about half of the variance observed. Our correlation tests show that the satellite spots and their local backgrounds are very strongly correlated, consistent with the correlated variations between the satellite spots primarily driven by the shared variations in the background levels. Once the shared variations are removed from the satellite spots, the lightcurves are dominated by only three principal components, one of which is accounting for about half of the total variance. These components and their modulations result in a remarkably stable anti-correlation pattern between the four spots over the four nights. }

The fact that spots are often anti-correlated may be attributable to phase errors, which may affect the flux balance between satellite spots. The fact that the anti-correlation pattern is repeating each night suggests that the noise factors -- perhaps phase errors -- that dominate the satellite spot variability also are repeatable. The repeatable structure in the satellite spots (about the same level of shared variations, the same anti-correlation pattern, the same number and relative contribution of principal components) suggests that with more satellite spot data a quantitative model could be constructed to robustly disentangle variations pertaining to individual spots or spot pairs (e.g. phase errors) from variations affecting all spots and targets (Strehl ratio changes or sky transparency variations).

\subsection{Optimal Photometry: SMAP}

By comparing several approaches to relative planet-to-planet photometry -- based on the standard deviation of magnitude differences between the four planets and four nights -- we were able to identify a procedure, SMAP, that represents a very significant improvement in the high-contrast photometric accuracy over standard methods. { Although satellite spots have been used recently for providing astrometric and photometric reference for measurements of exoplanets from Gemini Planet Imager observations \citep[][]{Galicher2014, Bonnefoy2014,Chilcote2015},  a unique and key element of our procedure is that we use the satellite spots as proxies of image quality on a frame-by-frame basis. By injecting artificial planets with brightness modulated on each frame by the mean satellite spot brightness (satellite spot-modulated artificial planet-injection photometry, SMAP) we partly correct for frame-to-frame variations in image quality and sky transparency.  Our detailed study of different reduction and photometric methods identified, as the most accurate combined reduction/photometry method, a combination of SMAP with a KLIP-based reduction applied to circular target areas centered on the planets, and parabola-fitted minimization of the planet-subtraction residuals. These components, combined, improve the typical relative planet-to-planet photometry from 0.25~mag to 0.1~mag or better.

We also point out that injecting more than 4 satellite spots can provide a better spatial sampling of phase errors and more measurements of the image quality variations across the field of view; therefore, this added information can further increase the accuracy with which the satellite spots can be used to correct for image quality. We suggest future tests with 8 or more satellite spots to establish a quantitative image quality correction model to further improve on the relative planet-to-planet photometry demonstrated here.}

\subsection{Relative H-band Planet Photometry in the HR~8799 system}

{ In Table~\ref{PublishedPhotometry} we compare our H-band relative photometry results with recently published high-quality measurements of the same four planets. All measurements shown in this table are relative to planet {\em b}, which is {\change the planet at the largest separation from the star}. We note that the measurements shown have all used slightly different H-band filters; nevertheless, given that the measurements are {\change relative and the planets have overall similar spectra in the H-band \citep[][]{Barman2011_HR8799,Oppenheimer2013,Pueyo2015,Barman2015,Bonnefoy2015}}, the color differences caused in the filter profile are expected to be small. Interestingly, although the uncertainty estimates for the individual measurements are large $(\sim0.14-0.18$~mag), {\change if the measurement in \citet[][]{Skemer2012} is excluded the} recent relative H-band {\em b-c} measurements agree within $\sim$0.01 magnitude. In contrast, although the photometric uncertainties are similar for the {\em b-d} pair, the scatter of the relative measurements for this pair is an order of magnitude higher than for {\change the {\em b-c} pair}. While the excellent agreement for the {\em b-c} planet pair in the time-averaged measurements is encouraging, this agreement is unlikely to exclude photometric variations that occur on timescales comparable to the integration times, as these would tend to average out in non-time-resolved observations. }

\begin{deluxetable}{lccccc}
\tabletypesize{\scriptsize}
\tablecaption{Relative H-band planet photometry from recent published studies. \label{PublishedPhotometry}}
\tablewidth{0pt}
\tablehead{
\colhead{Study}        & \colhead{Filter}        &  \colhead{Date}   & \colhead{b-c [mag]} & \colhead{b-d [mag]} & \colhead{b-e [mag]}    }
\startdata
Skemer et al. 2012   & LBT/Pisces H   &  10/16/2011      & 0.90$\pm$0.05 & 0.85 $\pm$0.2           & 1.2$\pm$0.2 \\
Esposito et al. 2013 & LBT/Pisces H  &   10/16/2011       &  1.00 $\pm $0.14&  0.72$\pm$0.19       & 1.37$\pm$0.44     \\
Zurlo et al. 2015  &  SPHERE H2  &     7/13/2014             &   0.99$\pm$0.18  & 1.06$\pm$ 0.22     &  1.17$\pm$0.24    \\   
Zurlo et al. 2015  &  SPHERE H3  &     7/13/2014             &   1.00$\pm$0.14  & 0.93$\pm$ 0.19   & 1.10 $\pm$0.23  \\   
This work & SPHERE Broad H  & 	Dec 2014              &   $1.01\pm$0.09      & 0.99$\pm$ 0.11     &   1.27$\pm$0.07 \\
\enddata
\end{deluxetable}

\subsection{Combining Relative Photometry from Different Nights}

Our study shows that relative planet-to-planet photometry is obtainable and comparable between different nights with a precision that is useful for exploring rotational modulations in planets, {\change which -- in brown dwarfs -- have observed amplitudes between 0.3\% and 27\%}. As the measurements are relative the data from different nights can be combined, which is important for two reasons.
First, for objects with limited visibility (such as the northern HR~8799 observed from the southern Cerro Paranal site) only few hour-long data sets can be readily obtained and
combining data from multiple nights may be the only option to build up the signal-to-noise required to confidently detect low-level (few percent amplitude) rotational modulations. Second, for slower rotators (Periods $>$6~h), sampling complete rotational phase curves will only be possible by combining data from multiple nights. 
As demonstrated here, the relative photometry possible in {\em multi-planet systems}, complemented with satellite-spot corrections, enables a self-consistent comparison of data from multiple nights and thus allows complete sampling of even slow rotators' lightcurves.

\subsection{Exoplanet Rotations and Variability with SPHERE}

The main purpose of this study is to identify the optimal approach for obtaining time-resolved relative planet photometry in high-contrast datasets and therefore our data is not intended to sample rotational modulations well. Nevertheless, for the completeness, we explore the upper limits that we can place based on our data sets on the rotational variability of the HR~8799 planets. Currently, the brown dwarf with the largest amplitude near-infrared variability is 2M2139 \citep[][]{Radigan2012,Apai2013,Buenzli2014b} with amplitudes in the J and H-bands of about 27\%
(the amplitude is thought to be time-variable due to light curve evolution in the object). Other brown dwarfs with high amplitude variability include SIMP0136 \citep[][]{Artigau2009,Apai2013} and Luhman 16 \citep[][]{Gillon2013,Biller2013,Burgasser2014,Buenzli2015}.
2M2139, the largest amplitude brown dwarf, has an approximately 8~h period; based on simple Rhines-scale arguments \citet[][]{Apai2013} argues that the largest atmospheric structures -- and thus highest rotational amplitudes --  are expected to occur in slow rotators, i.e. any object with similarly large amplitude variability will likely have periods longer than 4--6 hours. Assuming observations over three nights, each with 6 hours of continuous measurements and a cadence of 25 minutes, SPHERE would be able to obtain $\sim$40 measurements. With the peak-to-peak amplitude variations of 27\% seen in 2M2139 and a 40-point lightcurve with a photometric precision of 10\% or better, we estimate that the variations could be robustly distinguished  ($>15-20\sigma$) from a non-variable source, although we note that the actual significance level will depend on the rotational phases covered and the light curve shape. In an optimistic case, with both the minimum and maximum covered, even from observations covering one night the rotation period could be estimated.

Most sources will, of course, have smaller flux variations than 2M2139. These variations will be detected at correspondingly lower significance levels; however, as discussed in the previous section, observations from multiple nights can be combined to better constrain the light curves. Furthermore, as shown in \citet[][]{Heinze2015}, variability amplitudes in the red optical may be higher than those observed in the near-infrared, at least for late T-dwarfs. If this result is confirmed, SPHERE's and GPI's ability to work in the red optical wavelengths may be exploited in the exoplanet variability studies.

Lastly, we point out that our SPHERE Science Verification data was not only limited by the number of data points and relatively large airmasses during the observations, but also by the very small field rotations ($\sim8^\circ-9^\circ$). If more suitable exoplanet targets are found in the near future, larger field rotations should allow further improvement in the stellar PSF subtraction and, therefore, further improvement of the planet-to-planet photometric accuracy to levels below 10\%.

Thus, we conclude that SPHERE or GPI planet-to-planet photometry can reach an accuracy of 10\% or better with longer baseline observations, more favorable airmasses, and larger field rotations. {\change Even with the current precision using SMAP photometry it should be possible to measure the broad-band amplitudes and rotational periods of exoplanets from one or two nights of observations, if they resemble the properties of the most varying brown dwarfs.}

\section{Conclusions}
\label{Conclusions}

Using VLT/SPHERE/IRDIS we present here the first time-resolved observations of the four planets in the HR~8799~bcde exoplanet system, also including a detailed astrometric and photometric study, and tested and compared several approaches to derive accurate time-resolved photometry. Our key results are:\\
1) We determined the positions of the four planets for a new epoch (Dec 2015). \\
2) We measured the time-averaged relative brightness of the planets in H--band to an accuracy better than existing similar measurements.\\
3) We showed that using the normalized mean flux in the four satellite spots as a frame-by-frame flux correction for the injected artificial planets results in a significant improvement of the photometric accuracy.\\
4) We identified a reduction/photometry method that provides accurate time-resolved relative photometry in such a high-contrast case (satellite-spot-modulated artificial planet injection photometry, SMAP), reducing the planet-to-planet uncertainty from a standard aperture-based approach of 0.3~mag to below 0.1 mag.\\
5) We provided a detailed analysis of the behavior of the satellite spots and found that its nature is remarkably stable over the four nights and it is characterized by about 50\% shared variations and, on lower levels, by stable anti-correlation patterns between the spots. This result suggests that with more data a quantitative model could be constructed to accurately predict the intensity modulation
of the injected artificial planets as a function of their positions.\\
6) We demonstrated that in a multi-planet system it is possible to combine data from different nights and to reach relative planet-to-planet photometric accuracy of $\sim$10\% or better for 25 minute SPHERE blocks, even under non-optimal observing conditions (high airmasses, small field rotations, short integration times). It is very likely that with longer integration times and larger field rotations detecting variations typical of the large variations measured in some brown dwarfs would be feasible, and by combining data from multiple nights even slow rotators' phase curves can be fully mapped.

Therefore, we conclude that data sets spanning longer baselines in combination with satellite spot-modulated artificial planet-injection photometry (SMAP) will allow direct measurements of the rotation period and cloud cover in directly imaged exoplanets.

\acknowledgments
We thank the anonymous referee for the detailed, constructive, and prompt reviews which helped to improve the clarity of the manuscript. We are grateful for the SPHERE consortium for their hard work on building such a capable instrument, the ESO astronomers for carrying out the observations, and for the ESO User Support group for their assistance with the planning of the observations. E.B. was supported by the Swiss National Science Foundation (SNSF).
Based on observations made with ESO Telescopes at the La Silla Paranal Observatory under programme ID 60.A-9352(A). The results reported herein benefited from collaborations and/or information exchange within NASAÕs Nexus for Exoplanet System Science (NExSS) research coordination network sponsored by NASA's Science Mission Directorate.
{\it Facilities:} \facility{VLT (SPHERE)}.


\bibliographystyle{aa}       
\bibliography{apairefs}   

\end{document}